\documentclass[manuscript]{aastex}

\usepackage{color}
\usepackage{multicol}
\usepackage{graphicx}

\shorttitle{On the Validity of Collider-Mass Scaling}
\shortauthors{Walker et al.}

\begin{document}

\title{On the Validity of Collider-Mass Scaling for Molecular Rotational
Excitation}

\author{Kyle M. Walker, B. H. Yang, and P. C. Stancil}
\affil{Department of Physics and Astronomy and Center for Simulational Physics,
The University of Georgia, Athens, GA 30602, USA}

\author{N. Balakrishnan}
\affil{Department of Chemistry, University of Nevada, Las Vegas, NV 89154, USA}

\and

\author{R. C. Forrey}
\affil{Department of Physics, Pennsylvania State University, Berks Campus, 
Reading, PA 19610, USA}

\begin{abstract}

Rate coefficients for collisional processes
such as rotational and vibrational excitation are essential inputs in 
many astrophysical models. When rate coefficients are unknown, 
they are often estimated using known values from other systems. 
The most common example is to use He-collider rate coefficients to
estimate values for other colliders, 
typically H$_2$, using scaling arguments based on the reduced mass 
of the collision system.
This procedure is often justified by the assumption that the 
inelastic cross section is independent of the collider. 
Here we explore the validity of this approach
focusing on rotational inelastic transitions for collisions of
H, para-H$_2$, $^3$He, and $^4$He with CO in its vibrational ground state.
We compare rate coefficients obtained via explicit calculations to those deduced by
standard reduced-mass scaling.
Not surprisingly, inelastic cross sections and rate
coefficients are found to depend sensitively on both the
reduced mass and the interaction potential energy surface.
We demonstrate that standard reduced-mass scaling is
not valid on physical and mathematical grounds, and as a consequence, the common
approach of multiplying a rate coefficient for a molecule-He collision system
by the constant factor of $\sim$1.4 to estimate the rate coefficient 
for para-H$_2$ collisions is deemed unreliable. 
Furthermore, we test an alternative analytic scaling approach based on 
the strength of the interaction potential and the reduced mass of the collision systems.
Any scaling approach, however, may be problematic when low-energy resonances are
present; explicit calculations or measurements of rate coefficients
are to be preferred.

\end{abstract}


\section{Introduction}

The modeling of astrophysical observations requires a large
variety of fundamental data. In cool, low-density molecular
regions, collisional excitation/de-excitation rate coefficients are one class
of data needed for such models.
In many cases, however,  the (de)excitation rate coefficients 
for systems of astrophysical interest are not available.
While many molecular collisional calculations have been performed using
He as a collider due to its relative ease of computation, the 
dominant neutral species in many  
astrophysical environments is H$_{2}$ or H.
A common practice to obtain estimates for these unknown 
rates is to approximate them from known rate coefficients from
other collision systems, He colliders for example, by a reduced-mass scaling 
relation \citep[e.g.,][]{sch05,tak11}. 
This procedure has generally been ascribed to 
\citet{green78}, who predicted excitation rates of H$_{2}$ to be 
about 30\% higher than He rates in collisions with H$_{2}$CO, and later 
computed broadening cross sections from line-width 
parameters and predicted state-to-state excitation rate 
coefficients for CO-H$_{2}$O collisions \citep{green93}. 
These predictions were compared to theoretical He-CO rate coefficients and 
it was found that rate coefficients for excitation by water were related to 
those by He through the square root of the ratio of the systems' 
reduced masses.
Although the experimental data limited the applicability  of the predictions 
to room temperature and above, this ``standard" reduced-mass scaling 
relation has been used extensively for lower temperatures and 
for other collisional parameters, e.g., inelastic rate coefficients.

Recently, the accuracy of the standard reduced-mass
scaling approach has been tested for 
SiS \citep{liq08} and SO$_{2}$ \citep{cern11} excitation
and found to be inaccurate \citep[see also][]{rou13,tak11}.
In this paper, we revisit reduced-mass scaling 
and investigate how the reduced mass $\mu$ and interaction 
potential energy surface (PES) impact explicitly computed rate coefficients. 
Due to the wealth of data on CO, and its importance in astrophysics, it is
adopted here as a test molecule and used to
explore more physically reasonable analytical and scaling approximations 
to estimate unknown rate coefficients.

\section{Method}

Quantum mechanical close-coupling calculations were performed 
using the nonreactive scattering program MOLSCAT \citep{hut94} with the 
modern PESs of \citet{shep07}, \citet{jan05}, and 
\citet{hei97} for H-, H$_{2}$-, and He-CO, respectively.
The scattering cross sections were computed for kinetic energies between
10$^{-6}$ and 10$^3$~cm$^{-1}$ within the rigid-rotor approximation with
CO in its vibrational ground state and H$_2$ it is rovibrational ground state,
i.e. para-H$_2$.
Only para-H$_{2}$($j$=0) is considered in our study; 
ortho-H$_{2}$ and para-H$_{2}$($j$ \textgreater 0) rate coefficients obtained 
via any mass scaling approach from He data are unlikely to be accurate 
due to the lack of spherical symmetry of the internally excited molecule.
De-excitation rate coefficients as a function of temperature $T$
were obtained by thermally averaging the cross sections over a Maxwellian 
kinetic energy distribution given by
\begin{equation} \label{rate1}
k_{j\rightarrow j'}(T) = \left (\frac{8k_{B}T}{\pi \mu} \right )^{1/2}
\frac{1}{(k_{B}T)^{2}}\int_{0}^{\infty}\sigma_{j\rightarrow j'}(E_{k})
\exp(-E_{k}/k_{B}T)E_{k}dE_{k},
\end{equation}
where $\sigma_{j\rightarrow j'}(E_{k})$ is the 
state-to-state rotationally inelastic cross section,
$E_{k}$ the center of mass kinetic energy, 
$\mu = m_{X}m_{CO}/(m_{X} + m_{CO})$ the reduced mass 
of the $X$-CO system for collider $X$,
$k_{B}$ the Boltzmann constant, and $j$ the CO rotational quantum number.

\section{Results and Discussion}

\begin{figure}
\plotone{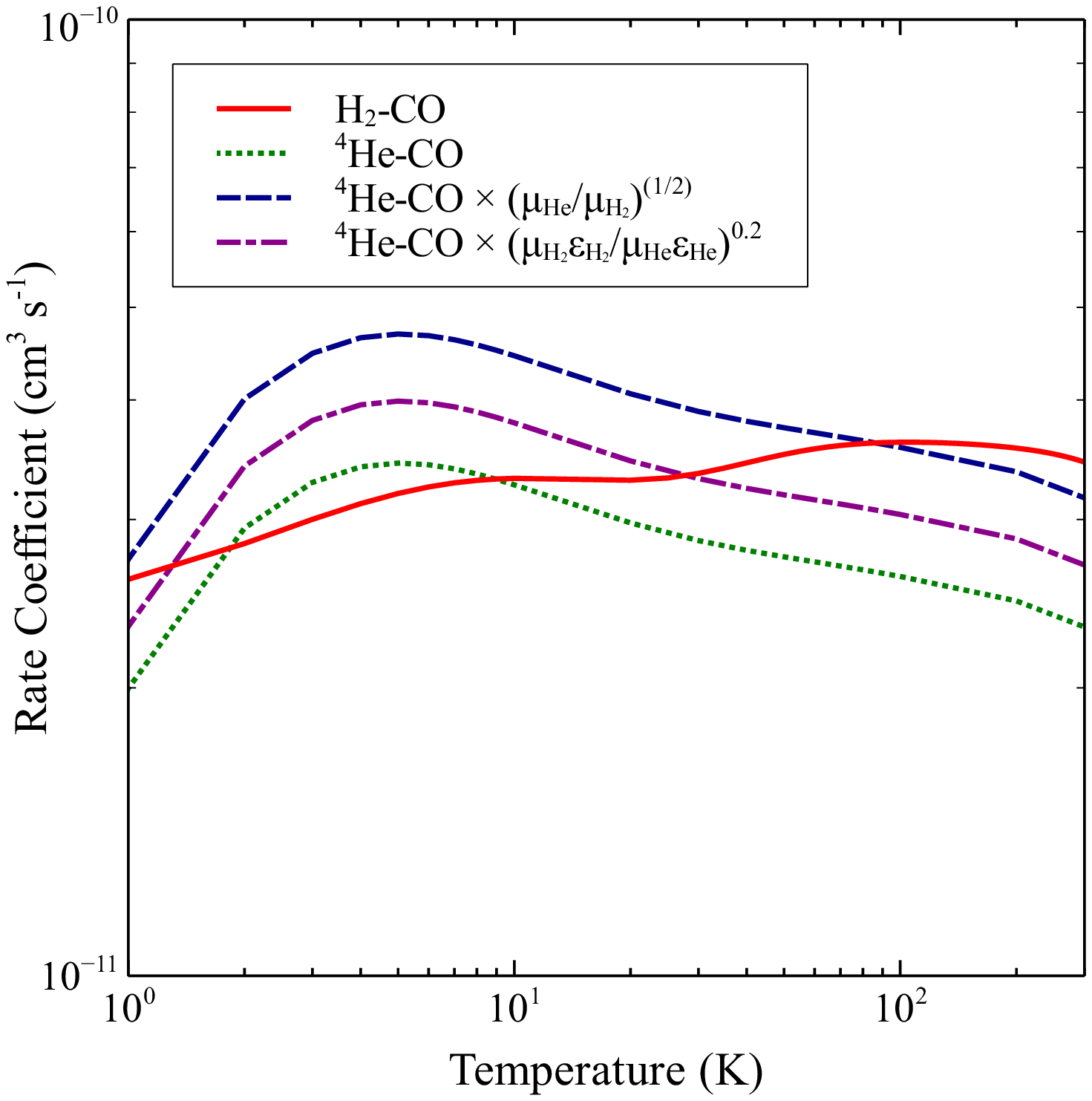}
\caption{Rate coefficients for the deexcitation of CO($j$=1) with the 
colliders He and para-H$_{2}$ and estimated values for H$_{2}$
via standard reduced-mass scaling and reduced-potential
scaling. 
\label{fig:figr1}}
\end{figure}

The main assumption of standard reduced-mass scaling is 
the statement that the cross section as a function of $E_k$ is 
independent of the collider. It is then argued
that the rate coefficients between He-CO and 
H$_{2}$-CO, for example, scale as the square root of the ratio of reduced 
masses according to
\begin{equation} \label{scale}
k^{\rm H_2}_{j\rightarrow j'}(T) = 
\left (\frac{\mu_{\rm He}}{\mu_{\rm {H_{2}}}}\right)^{1/2}
k^{\rm He}_{j\rightarrow j'}(T),
\end{equation}
as deduced from the prefactor of Equation~(\ref{rate1}). Here
$k^{X}_{j\rightarrow j'}(T)$ is the state-to-state rate coefficient for collider $X$, 
$\mu_{X}$ is the reduced mass for the $X$-CO system, and
a prefactor of $\sim$1.4 is obtained in this case. As an illustration,
Figure~\ref{fig:figr1} displays the calculated rate coefficients for the 
deexcitation of CO($j$=1) with the colliders He and H$_{2}$. 
The estimated H$_{2}$ rate coefficients using the standard reduced-mass 
scaling relation, Equation~(\ref{scale}), are seen to 
deviate significantly from the explicitly 
computed values, especially for $T \lesssim 100$~K.

Although the standard reduced-mass scaling relation has been widely 
adopted \citep[e.g.,][]{sch05,ada13,mat14}, it actually assumes 
not that the cross section $\sigma(E_{k})$, but 
that the integral in Equation~(\ref{rate1}), 
is independent of the collider.
However, this assumption is not generally valid because: (a) the 
cross section depends on the adopted PES and (b) the 
kinetic energy depends on $\mu$.
In cases where Equation~(\ref{scale}) 
has produced reasonable estimates, 
it may have been the result of fortuitous 
cancelation of the effects due to points (a) and (b).

\begin{figure}
\plotone{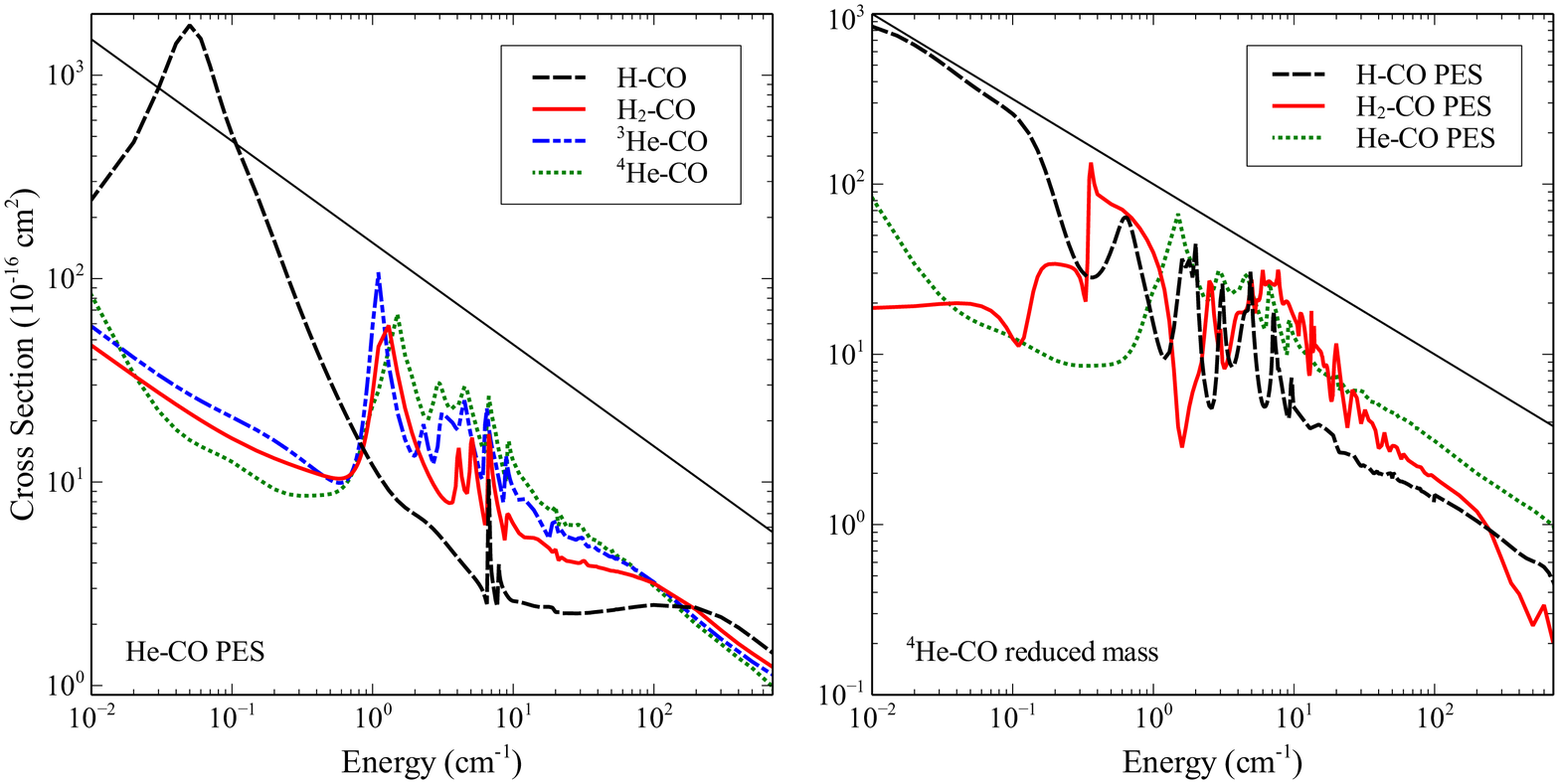}
\caption{(a)~Cross sections of the colliders 
H, para-H$_{2}$, $^{3}$He, and $^{4}$He 
on the He-CO PES for the $j=1\rightarrow 0$ transition. 
(b)~Cross sections for the same transition using the 
$^{4}$He-CO reduced mass on the H-, H$_{2}$-, and He-CO PESs.
The straight solid line indicates 
a $1/v$ cross section dependence.
\label{fig:figr2}}
\end{figure}

To explicitly illustrate the failings of Equation~(\ref{scale}) and to
explore other more physically-motivated approaches, we investigate 
three tracks: i) the behavior of inelastic cross sections as a function of system parameters,
ii) prediction of $k_{j\rightarrow j'}(T)$ adopting analytical relations 
for the cross section,
and iii) an alternative scaling approach based on the well-depth of the PES.  
To test the dependence of the cross section on $\mu$ and the PES, a series of 
calculations were performed for the collider masses H, H$_{2}$, $^{3}$He, 
and $^{4}$He with CO on each of the H-CO, H$_{2}$-CO, and He-CO 
PESs with some examples shown in Figure~\ref{fig:figr2}. 
In Figure~\ref{fig:figr2}a, the cross sections, which 
were all calculated on the He-CO PES, are seen
to depend significantly on the adopted $\mu$. In addition to
changes in the cross section slopes and magnitudes, the positions
of quasibound resonances vary, especially for H$_{2}$ and He compared
to H.
Figure~\ref{fig:figr2}b shows the results of calculations using
the $^{4}$He-CO reduced mass on the three different PESs.
Results for the other three masses (not shown) were likewise 
found to be sensitive to the PES.
As expected, the cross section does indeed
depend on both the PES and $\mu$, and the assumptions implicit in
Equation~(\ref{scale}) are not valid.

As a yet further illustration, Equation~(\ref{rate1}) 
can be rewritten with the cross section given in terms of the relative velocity
$v$ of the collision system
\citep{flo90};
\begin{equation} \label{rate2}
k_{j\rightarrow j'}(T) =  \left (\frac{2}{\pi} \right)^{1/2} 
\left (\frac{\mu}{k_{B}T} \right)^{3/2}
\int_{0}^{\infty}\sigma_{j\rightarrow j'}(v)
\exp(-\mu v^{2}/2k_{B}T)v^{3}dv.
\end{equation}
This leaves $\mu$ in both the exponential 
Boltzmann term in the integral and in the prefactor and 
shows that the original arguments justifying 
standard reduced-mass scaling should be reconsidered
(compare to Equation~\ref{rate1}). To gain 
additional insight and to explore an alternate scaling approach, 
assume the cross section to have the
analytical form
\begin{equation} \label{sigma}
\sigma_{j\rightarrow j'}(v) = Bv^{a},
\end{equation}
where $B$ is an (undetermined) constant and $a$ is some power.
This leads to rate coefficients of the form \citep{stan98}
\begin{equation} \label{krate}
k(T)=A(a)B(T/\mu)^{b},
\end{equation}
where $b=(1/2)(a+1)$ and $A$ is a function of $a$, both deduced from the
Gaussian integral in Equation~(\ref{rate2}). This result is exact, given the assumption
of Equation~(\ref{sigma}), and applicable to all collision systems. Therefore, if
$\sigma(v)$ is assumed to be independent of the collider as originally 
supposed, Equation~(\ref{scale}) is corrected by replacing the square-root with the 
exponent $b$ yielding
\begin{equation} \label{scale2}
k^{\rm Z}_{j\rightarrow j'}(T) = 
\left (\frac{\mu_{\rm Y}}{\mu_{\rm {Z}}}\right)^{b}
k^{\rm Y}_{j\rightarrow j'}(T).
\end{equation}
The scaling equation is now general for any two colliders Y and Z and any
dependence of cross section on energy.
Only for a constant cross section will $b$ equal 1/2.

\begin{figure}
\plotone{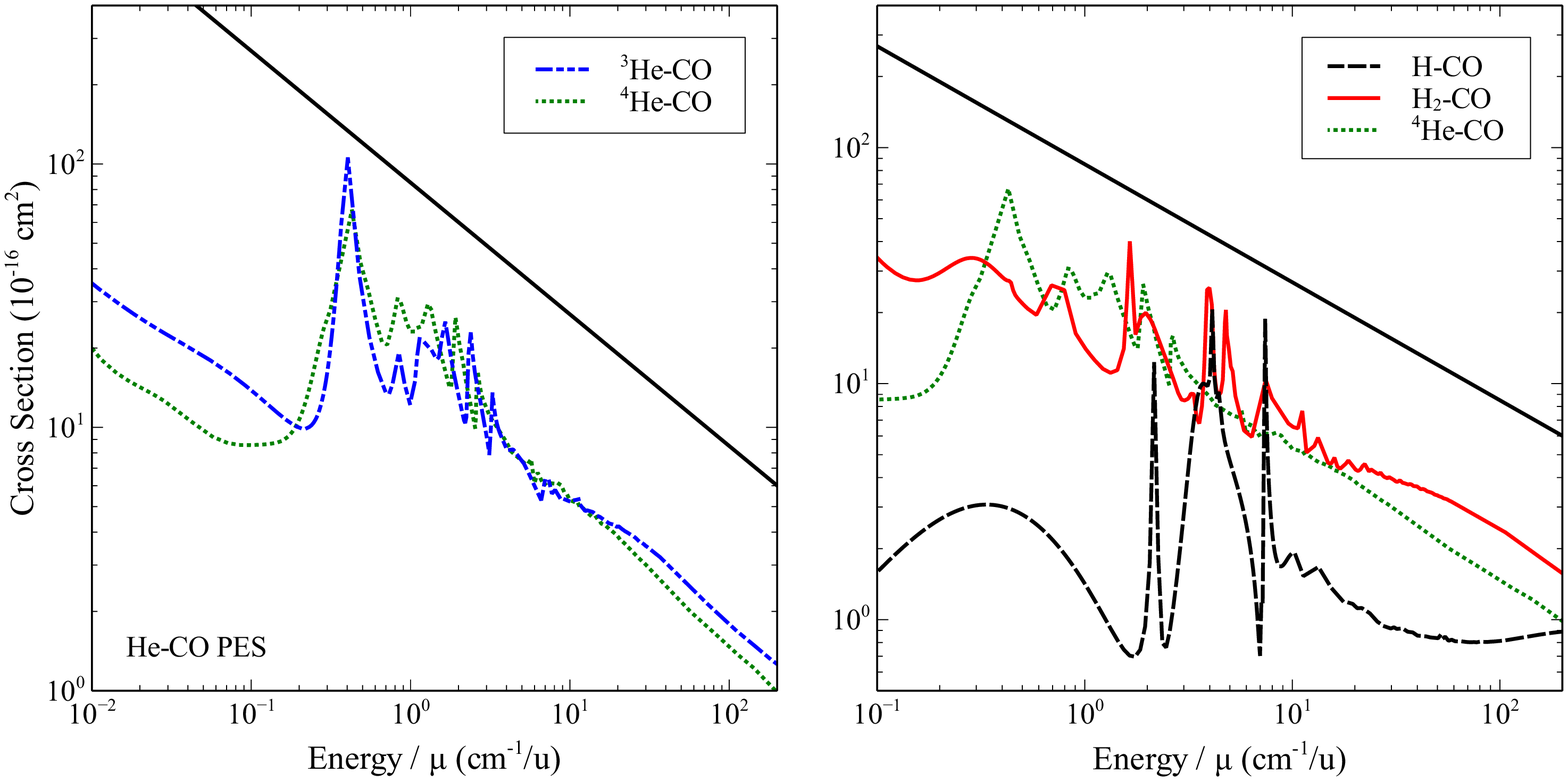}
\caption{Cross sections for the $j=1\rightarrow 0$ transition 
as a function of kinetic energy/$\mu$ for
(a)~the colliders $^{3}$He and $^{4}$He on the He-CO PES. 
(b)~Calculated cross sections on their respective PESs. 
The straight solid line indicates 
a $1/v$ cross section dependence.
\label{fig:figr3}}
\end{figure}

Figure~\ref{fig:figr3}a displays the cross section dependence for 
$^{3}$He-CO and $^{4}$He-CO, but plotted as a function of the
kinetic energy divided by $\mu$, which is proportional
to $v^2$.
Above $\sim$0.1~cm$^{-1}$u$^{-1}$, the cross section is relatively 
independent of $\mu$, a concept well-known in 
ion-atom collisions \citep[e.g.,][]{stan95}.
Figure~\ref{fig:figr3}b gives a related plot where He and 
para-H$_2$ colliders give qualitatively
similar behavior, with the background cross sections of each falling-off with a $1/v$
dependence for $E\gtrsim 2$ cm$^{-1}$u$^{-1}$.
Combining these observations with Equations~(\ref{krate}) 
and (\ref{scale2}), the resulting
rate coefficients, neglecting the resonances, will be relatively independent
of both $T$ and $\mu$ (i.e., $a=-1$, $b=0$).
The H cross section is smaller due to a considerably different PES structure
\citep[see][]{shep07}.

\begin{figure}
\plotone{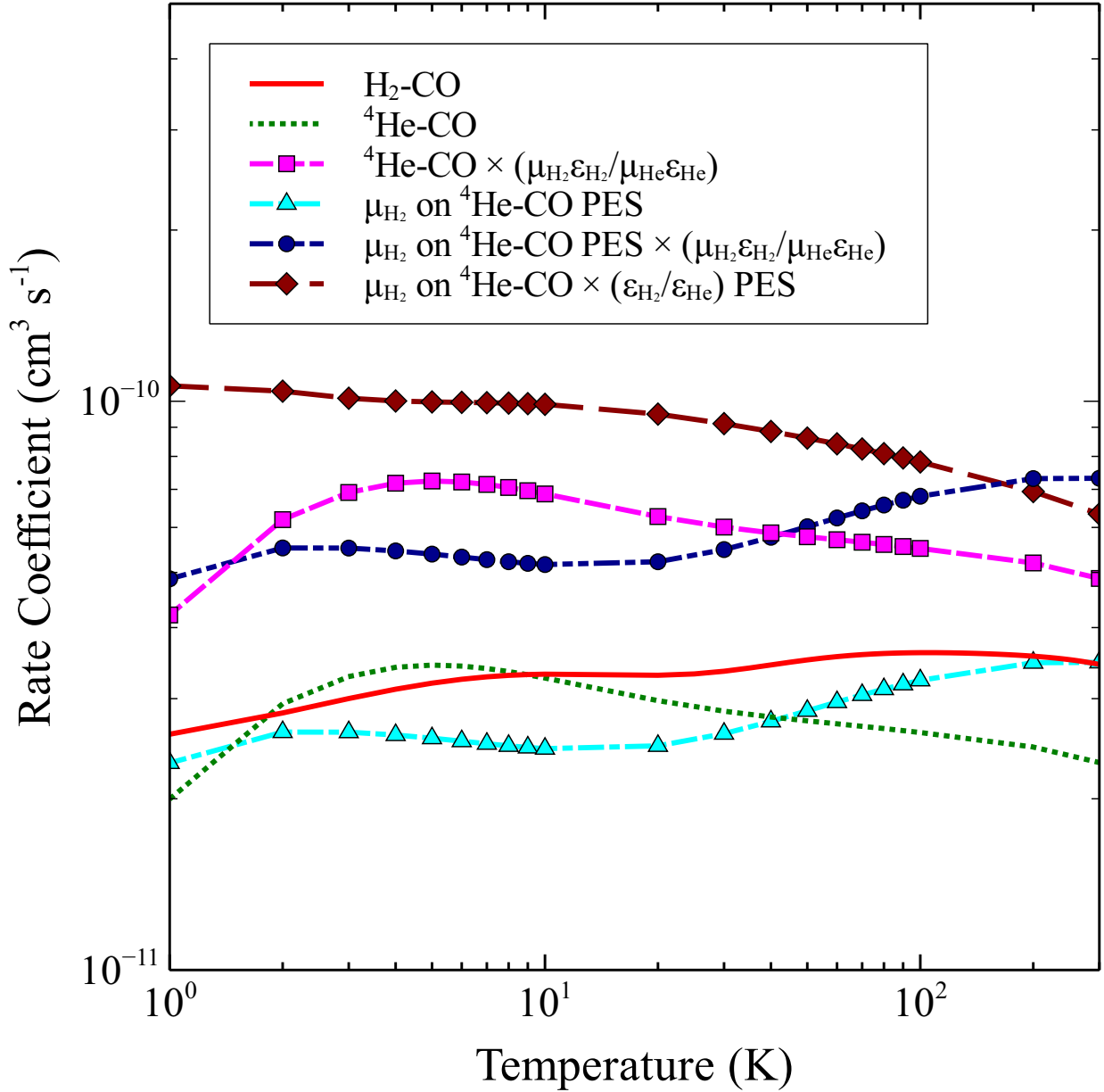}
\caption{Rate coefficients for the deexcitation of CO($j$=1) with the 
colliders He and para-H$_{2}$ and estimated values for H$_{2}$
via possible scaling relations. See text for discussion.
$\varepsilon$$_{\rm{H_{2}}}=93.1$~cm$^{-1}$ \citep{jan05} and 
$\varepsilon$$_{\rm{He}}=23.7$~cm$^{-1}$ \citep{hei97}.
\label{fig:figr4}}
\end{figure}

As possible intermediate methods between scaling and explicit
calculations, we attempted four other approaches to obtain
predictions for para-H$_{2}$ as depicted in Figure~\ref{fig:figr4}a. 
First, considering the differing PESs, a possible scaling  
is obtained from the ratio of the reduced potentials $\mu_{X}\varepsilon_{X}$
\citep{joa79},
where $\varepsilon_{X}$ is the van der Waals well-depth of the PES.
However, the H$_{2}$ rate coefficients for
the $1\rightarrow 0$ transition are overestimated (see below).
Second, explicit scattering calculations using 
$\mu_{\rm H_2}$ on the He-CO PES gave 
rates in reasonable agreement with the explicitly 
calculated H$_{2}$-CO results.
Third, multiplying these rates by the ratio of the well depths alone 
(a factor of $\sim$3.9, not shown), and by the ratio of reduced
potentials again overestimated the H$_{2}$ rate coefficients.
Finally, cross sections using $\mu_{\rm H_2}$
on the He-CO PES scaled to match $\varepsilon_{\rm H_2}$ were computed, 
but the resulting rate coefficients
overestimated the explicit H$_2$-CO rates. 
Of these, the second approach, which used the simpler 
2D He-CO PES with $\mu_{\rm H_2}$ appears to give the best results, 
but still requires new scattering calculations.

Considering the above findings, we arrive at the two most promising scaling
options. In the first case,
Equation~(\ref{scale2}) can be applied above
$\sim$10-50~K when $k_{j\rightarrow j'}(T)$ is known for He. 
The lower limit can be estimated with knowledge
of $\varepsilon_{X}$ which is roughly equal to the upper kinetic energy limit
of the quasibound resonances (see Figs.~\ref{fig:figr2}~-~\ref{fig:figr3}). 
For example, if $b=0$ ($a=-1$), the rate coefficients are independent 
of $T$ and $\mu_{X}$, as opposed to Equation~(\ref{scale}). 
Of course $B$ is assumed to be the same for He and H$_2$ colliders.

\begin{figure}
\plotone{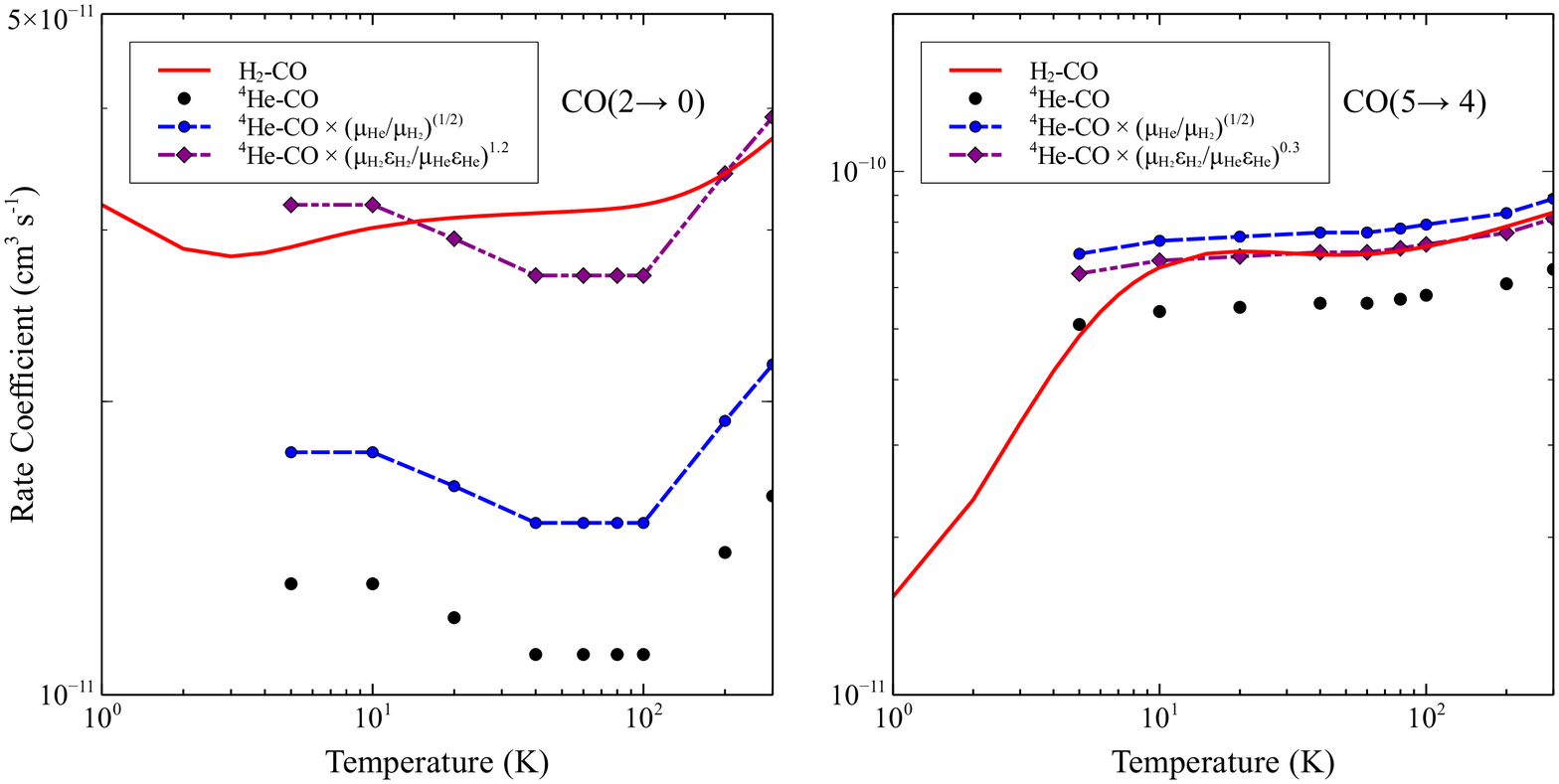}
\caption{
Rate coefficients for the (a)~$j=2\rightarrow 0$ and (b)~$j=5\rightarrow 4$
transitions of CO with H$_{2}$ \citep{yang10} and 
He \citep{cecchi02} compared to the predictions of 
standard reduced-mass scaling and reduced-potential scaling with $C=1.2$ 
and $0.3$, respectively.
\label{fig:figr5}}
\end{figure}

The second option is appropriate for $T\lesssim 10-100$~K 
where rate coefficients are highly sensitive to quasibound resonances.
These resonances may partially be accounted for by scaling via the ratio 
of the reduced potentials
$\mu_{X}\varepsilon_{X}$, as discussed above, with a phenomenological exponent $C$
according to 
\begin{equation} \label{scale3}
k^{\rm Z}_{j\rightarrow j'}(T) = 
\left (\frac{\mu_{\rm Z}\varepsilon_{\rm Z}}{\mu_{\rm Y}\varepsilon_{\rm Y}}\right)^{C}
k^{\rm Y}_{j\rightarrow j'}(T).
\end{equation}
Numerical values for the van der Waals well-depths of the interaction PESs are 
generally available from experimental and theoretical work in the chemical 
physics community \citep[see][]{rad80}.
This scaling option was explicitly tested for fifteen 
$\Delta j$ transitions using He-CO \citep{cecchi02} and H$_2$-CO
\citep{yang10} theoretical data.
The exponent $C$ was optimized to minimize the scaling residuals from 
5$\sim$500~K.
Figure~\ref{fig:figr5}a,b shows that both the 2$\rightarrow$0 and 5$\rightarrow$4 
rate coefficients scaled via the reduced potential method give the best estimates.
In fact, even $\Delta j$ transitions scaled by the reduced potential ratio with 
exponent $C\sim 0.7-1.3$ give good predictions for H$_2$-CO rate coefficients.

Although standard reduced-mass scaling reproduces the H$_2$-CO data for
odd $\Delta j$ transitions more accurately than even $\Delta j$ transitions,
the agreement is fortuitous..
Reduced-potential scaling, on the other hand, with $C\sim 0.0-0.4$ for 
odd $\Delta j$ transitions, shows improvement of the predictions \textit{and}
is based on the physical properties of the interacting system.
We note that near-homonuclear molecules, such as CO,
follow propensity rules whereby odd $\Delta j$ transitions are suppressed 
compared to even $\Delta j$ transitions, and it seems the dichotomy of the 
phenomenological exponent $C$ expresses this propensity.

To determine the accuracy of the new reduced-potential scaling approach, 
we calculated the normalized root-mean-square deviation(NRMSD), $\sigma_{norm}$, 
of the H$_2$ rate coefficient predictions for both standard 
reduced-mass scaling and reduced-potential scaling, given by:
\begin{equation} \label{nrmsd}
\sigma_{norm} =
\frac{\sqrt{\frac{\sum\limits_{T=i}^{N} (k_{scale}(T)-k_{calc}(T))^2}{N}}}
{k_{max} - k_{min}},
\end{equation}
where $N$ is the number of temperature data points and $k_{max}$ and $k_{min}$ are the 
values of the maximum and minimum rate coefficients, respectively.
The resulting percentage indicates the residual variance between the calculated 
H$_2$ rate coefficients, $k_{calc}$, and those scaled from He, $k_{scale}$.
Table~\ref{table1} lists these values for fifteen transitions of CO.
There is a remarkable improvement in reduced-potential scaling predictions 
for even $\Delta j$ transitions. Odd $\Delta j$ transitions also show
improved predictions of reduced-potential scaling over standard reduced-mass 
scaling, albeit less so.
These odd $\Delta j$ transitions exhibit the broadest range in rate coefficients 
and can vary more than an order of magnitude across the temperature range 2-500~K,
whereas rate coefficients for the even $\Delta j$ transitions are primarily 
flat across this range. 
Hence the odd $\Delta j$ transitions contain a larger residual variance. 
 
\begin{figure}
\plotone{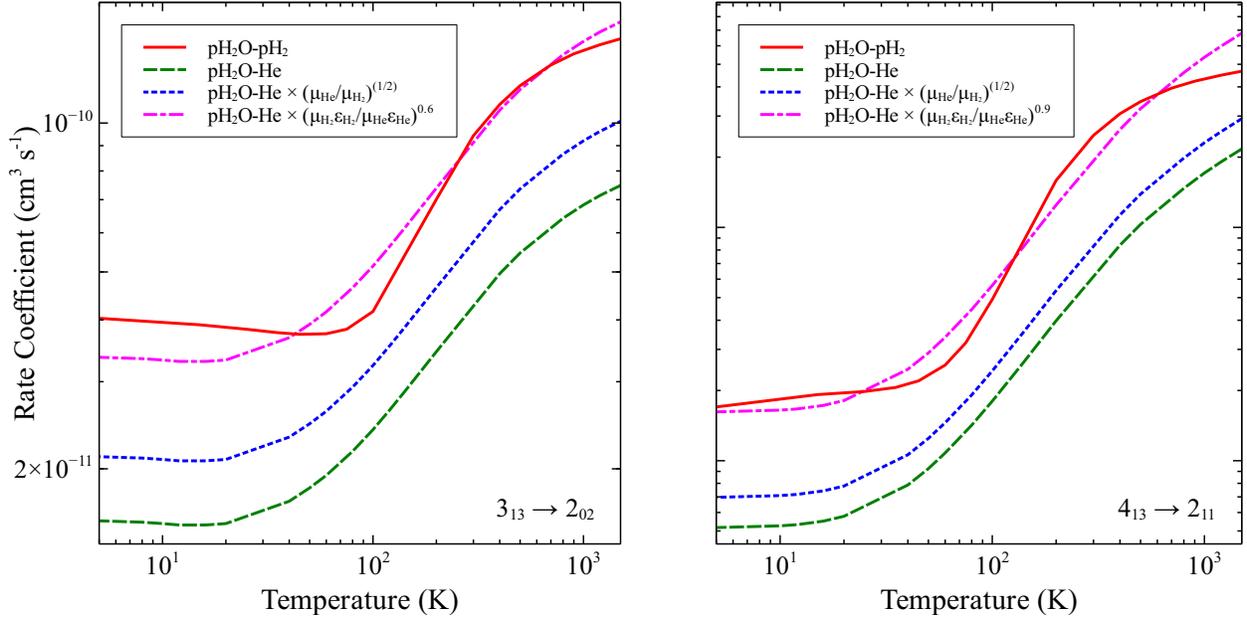}
\caption{
Rate coefficients for the deexcitation of 
H$_2$O($j_{k_{a}k_{c}}$), (a)~$3_{31} \rightarrow 2_{20}$, 
(b)~$4_{13} \rightarrow 2_{11}$, with 
para-H$_2$ \citep{dub09} and He \citep{yang13a} compared to standard reduced-mass 
scaling and reduced-potential scaling with $C=0.6$ and 0.9, respectively. 
The water well depths are
$\varepsilon_{\rm He}=34.4$~cm$^{-1}$ \citep{pat02} and 
$\varepsilon_{\rm H_2}=221.9$~cm$^{-1}$ \citep{fau05}.
\label{fig:figr6}}
\end{figure}

Figure~\ref{fig:figr6}a,b gives an example of a similar study 
of reduced-potential and standard reduced-mass scaling for H$_2$O to due He and 
para-H$_2$ collisions. 
From a survey of 32 transitions, the dominant transitions which obey the 
propensity rules
$|\Delta j| = |\Delta k_a| = |\Delta k_c| = 1$
are reasonably reproduced by the 
reduced-potential approach with $C\sim 0.6-0.8$, 
while the subdominant transitions 
$|\Delta j| = 1$, $\Delta k_a = 0$, $\Delta k_c =\pm 2$ or 
$\Delta k_a = \pm 2$, $\Delta k_c = 0$ and
$|\Delta j| = 2$, $\Delta k_a = 0$, $\Delta k_c =\pm 2$ or 
$\Delta k_a = \pm 2$, $\Delta k_c = 0$ 
extend this range to $C\sim 0.5-1.2$.
Cases which are reproduced best with $C \lesssim 0.5$ or $\gtrsim 1.2$ 
typically correspond to weak transitions with rate coefficients 2-3 orders
of magnitude smaller than the dominant transitions so that errors in their 
prediction are of less significance. 

\begin{figure}
\plotone{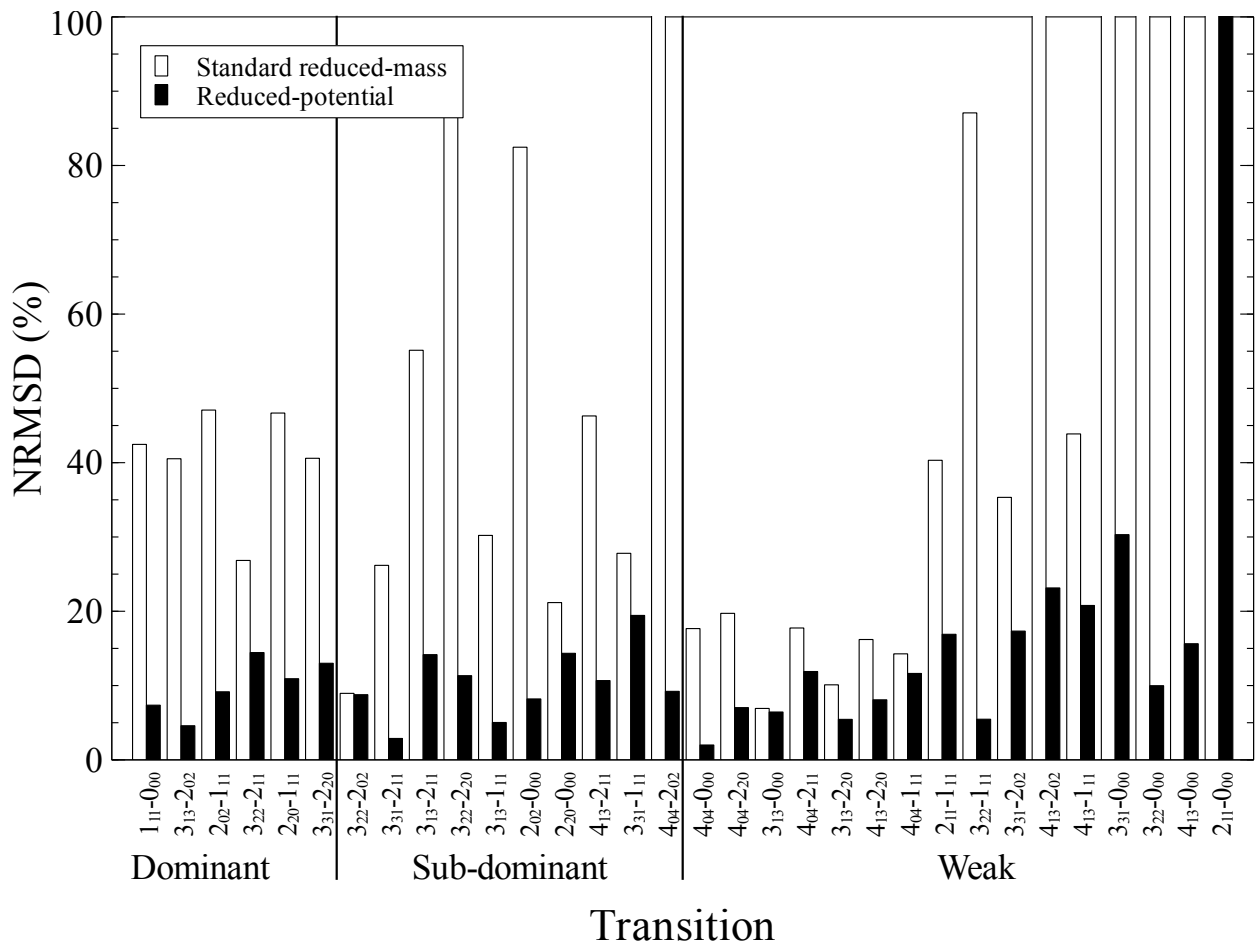}
\caption{
Normalized root-mean-square deviation(NRMSD) in standard reduced-mass 
scaling and reduced-potential scaling for 32 transitions of H$_2$O, truncated 
at 100$\%$. The dominant, sub-dominant, and weak transitions are further 
organized from left to right in increasing values of the exponent $C$.
\label{fig:figr7}}
\end{figure}

The reliability of the reduced-potential scaling method was again addressed by 
computing the NRMSD and comparing it to the NRMSD of standard reduced-mass 
scaling.  
Figure~\ref{fig:figr7} compares the NRMSD for both methods for each transition of 
H$_2$O.
Predictions from reduced-potential scaling exhibit less residual variance 
in all 32 transitions, with a mean of 35$\%$ or less, and clearly demonstrate the 
superiority of the new reduced-potential scaling method.

\begin{figure}
\epsscale{1.0}
\plotone{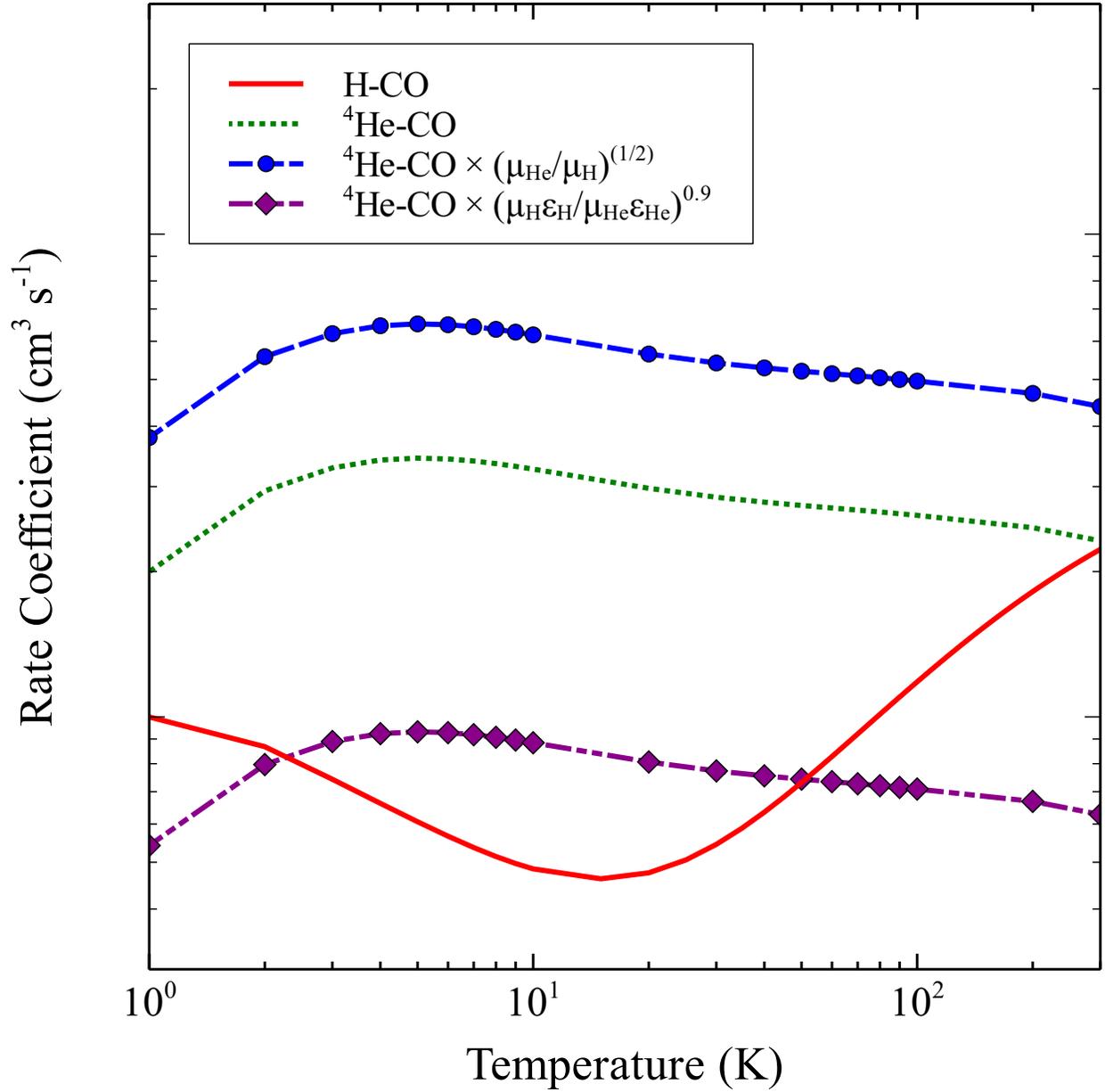}
\caption{
Rate coefficients for the deexcitation of CO($j$=1) with 
H \citep{yang13} and He compared to standard reduced-mass 
scaling and reduced-potential scaling with $C=0.9$.
\label{fig:figr8}}
\end{figure}

Finally, Figure~\ref{fig:figr8} shows the two scaling methods for H-CO collisions, 
where the reduced-potential with $C = 0.9$ rather than standard 
reduced-mass scaling more accurately predicts the calculated rate 
coefficients.
While additional studies of reduced potential scaling on a variety of other molecules are
needed and in progress, the cases studied here suggest that the approach can reasonably
predict rate coefficients for dominant transitions with $C\sim 0.8$, while $C<0.4$ can 
account for weak transitions with the partitioning predicted from known 
propensity rules.

\section{Conclusion}

Rotational inelastic transitions for 
collisions of H, para-H$_{2}$, $^{3}$He, and $^{4}$He with CO($j$=1) 
using three PESs were computed to study the 
cross section dependence on reduced mass $\mu$ and interaction potential  
with the goal of gaining insight into rate coefficient scaling.
Although earlier investigations indicated that scaling via the 
ratio of the square root of reduced masses gave reasonable 
estimates for collisional rate coefficients, the current study, 
shows that this agreement was 
fortuitous \citep[see also][for similar findings for HD]{sch90}.
The constant factor of $\sim$1.4 frequently used to predict the rate 
coefficients of para-H$_2$ from that of He generally lead to inaccurate 
results due to the fact that the underlying assumptions are not valid.
Scaling by this standard reduced-mass relation is therefore not 
recommended. Two alternative scaling approaches are proposed.
In the first case, if the inelastic cross section can be represented by
an analytical function of the relative velocity, then an exact rate
coefficient scaling exists as a function of $T$ and $\mu$,
valid for all collision systems. A second
approach, which accounts for the contribution of low-energy quasibound
resonances, is based on ratios of the product of $\mu$ with the PES
well-depth. Preliminary testing of the reduced-potential method 
in conjunction with known propensity rules for CO and H$_2$O gives 
reasonable predictions.
While these two approaches may lead to mathematically
and physically-reasonable scalings, it is only through explicit calculation
and/or measurements that reliable inelastic rate coefficients can
be obtained. 
The improved scaling approaches proposed here 
may provide useful estimates until
such explicit data become available.

\acknowledgments

We thank Peter van Hoof, Floris van der Tak, John Black, Gary Ferland 
and Ryan Porter for helpful discussions. This work was partially 
supported by NASA grant NNX12AF42G and a grant from the UGA 
Provost’s Office.

\clearpage

\clearpage

\clearpage

\clearpage

\clearpage

\clearpage

\clearpage

\begin{deluxetable}{rrccc}
\tablecolumns{5}
\tablewidth{0pc}
\tablecaption{The optimized values of $C$ and their respective normalized 
root mean square deviations (NRMSDs) for collisional deexcitation transitions 
of CO with H$_{2}$ and He scaled via the standard reduced-mass(rm) and new
reduced-potential(rp) methods.
\label{table1}}
\tablehead{
\colhead{$\Delta j$}    &   \colhead{$j\rightarrow j'$}   &
\colhead{C} & \colhead{NRMSD$_{\rm rm}$}   & \colhead{NRMSD$_{\rm rp}$}
}
\startdata
Even        &	2-0	&	1.2	&	138.60	&	17.95	\\
\nodata     &	3-1	&	1.2	&	125.31	&	18.04	\\
\nodata	&	4-2	&	1.2	&	133.00	&	17.64	\\
\nodata	&	5-3	&	1.3	&	141.07	&	25.73	\\
\nodata	&	4-0	&	0.9	&	50.16	&	9.06	\\
\nodata	&	5-1	&	0.7	&	28.57	&	4.07	\\
\hline
Odd	&	4-3	&	0.4	&	27.41	&	26.63	\\
\nodata	&	3-2	&	0.4	&	34.38	&	34.99	\\
\nodata	&	2-1	&	0.4	&	40.07	&	40.89	\\
\nodata	&	5-4	&	0.3	&	31.62	&	18.81	\\
\nodata     &	1-0	&	0.3	&	52.43	&	41.02	\\
\nodata	&	5-0	&	0.1	&	20.06	&	13.79	\\
\nodata	&	4-1	&	0.1	&	35.88	&	22.23	\\
\nodata	&	3-0	&	-0.2	&	38.91	&	29.48	\\
\nodata	&	5-2	&	-0.2	&	38.59	&	34.45	\\
\enddata
\end{deluxetable}


\begin{thebibliography}{}

\bibitem[Adande, Edwards, \& Ziurys(2013)]{ada13} Adande, G. R., Edwards, 
    J. L., \& Ziurys, L. M.
    2013, \apj, 778, 22

\bibitem[Cecchi-Pestellini et al.(2002)]{cecchi02} Cecchi-Pestellini, C.,
    Bodo, E., Balakrishnan, N., \& Dalgarno, A. 
    2002, \apj, 571, 1015

\bibitem[Cernicharo et al.(2011)]{cern11} Cernicharo, J., et al.
    2011, \aap, 531, 103

\bibitem[Dubernet et al.(2009)]{dub09} Dubernet, M.-L., Daniel, F., Grosjean, A., 
    \& Lin, C. Y. 
    2009, \aap, 497, 911 

\bibitem[Faure et al.(2005)]{fau05} Faure, A., et al. 
    2005, J. Chem. Phys., 122, 221102

\bibitem[Flower(1990)]{flo90} Flower, D. 1990, Molecular Collisions
    in the Interstellar Medium (Cambridge: Cambridge Univ. Press)

\bibitem[Green(1993)]{green93} Green, S. 
    1993, \apj, 412, 436

\bibitem[Green et al.(1978)]{green78} Green, S., Garrison, B. J., Lester, W. A., \& 
    Miller, W. H.
    1978, \apjs, 37, 321

\bibitem[Heijmen et al.(1997)]{hei97} Heijmen, T. G. A., Moszynski, R.,
    Wormer, P. E. S., \& van der Avoird, A. 
    1997, J. Chem. Phys., 107, 9921

\bibitem[Hutson \& Green(1994)]{hut94} Hutson, J. M. \& Green, S. 1994,
    MOLSCAT Computer Code (Version 14, Distributed by Collaborative
    Computational Project No. 6; Swindon, UK: Engin. \& Phys. Sci.
    Res. Council)

\bibitem[Jankowski \& Szalewicz(2005)]{jan05} Jankowski, P. \& Szalewicz, K. 
    2005, J. Chem. Phys., 123, 104301
    
\bibitem[Joachain(1979)]{joa79} Joachain, C. J. 1979, Quantum Collision Theory
    (Amsterdam: North-Holland)

\bibitem[Lique et al.(2008)]{liq08} Lique, F., et al.
    2008, \aap, 478, 567

\bibitem[Matsuura et al.(2014)]{mat14} Matsuura, M. et al.
    2014, \mnras, 437, 532

\bibitem[Patkowski et al.(2002)]{pat02} Patkowski, K., et al.
    2002, J. Mol. Struct., 591, 231

\bibitem[Radzig \& Smirnov(1980)]{rad80} Radzig, A. A., \& Smirnov, B. M. 
    1980, Reference Data on Atoms, Molecules, and Ions (Berlin: Springer)

\bibitem[Roueff \& Lique(2013)]{rou13} Roueff, E. \& Lique, F. 
    2013, Chem. Rev., 113, 8906

\bibitem[Schaefer(1990)]{sch90} Schaefer, J.
    1990, \aaps, 85, 1101

\bibitem[Sch\"oier et al.(2005)]{sch05} Sch\"oier, F. L., van der Tak, F. F. S., 
   van Dishoeck, E. F., \& Black, J. H. 
   2005, \aap, 432, 369

\bibitem[Shepler et al.(2007)]{shep07} Shepler, B. C., et al.
    2007, \aap, 475, L15

\bibitem[Stancil et al.(1998)]{stan98} Stancil, P. C., Lepp, S.,
    \& Dalgarno, A. 1998, \apj, 509, 1

\bibitem[Stancil \& Zygelman(1995)]{stan95} Stancil, P. C.
    \& Zygelman, B. 1995, \prl, 75, 1495

\bibitem[Van der Tak(2011)]{tak11} Van der Tak, F. F. S. 2011, in 
    IAU Symposium 280, The Molecular Universe, ed. J. Cernicharo \&
     R. Bachiller (Cambridge: Cambridge Univ. Press) 449

\bibitem[Yang et al.(2013a)]{yang13a} Yang, B. H., et al. 
    2013a, \apj, 765, 77

\bibitem[Yang et al.(2010)]{yang10} Yang, B. H., Stancil, P. C., Balakrishnan, N. \&
    Forrey, R. C. 
    2010, \apj, 718, 1062

\bibitem[Yang et al.(2013b)]{yang13} Yang, B. H., et al. 
    2013b, \apj, 771, 49

\end{thebibliography}
\end{document}